\documentclass[prd,twocolumn,reprint,preprintnumbers,nofootinbib,superscriptaddress]{revtex4-1}
\pdfoutput=1
\usepackage{cancel}

\RequirePackage[colorlinks=true
,urlcolor=blue
,anchorcolor=blue
,citecolor=blue
,filecolor=blue
,linkcolor=blue
,menucolor=blue
,linktocpage=true
,pdfproducer=medialab
,pdfa=true
]{hyperref}

\usepackage{bbold}
\usepackage{amsmath,amssymb,amsthm,amsfonts}
\usepackage{graphicx,tabularx}
\usepackage{color}
\usepackage{multirow}
\usepackage{comment}
\usepackage{enumitem}
\usepackage{cleveref}
\usepackage{slashed}
\usepackage[normalem]{ulem}
\usepackage{scalerel}
\setlist[description]{leftmargin=0.3cm}
\setlist[itemize]{leftmargin=0.5cm}

\newcommand{\be}{\begin{equation} \begin{aligned}}
\newcommand{\ee}{\end{aligned} \end{equation}}
\newcommand{\beqa}{\begin{eqnarray}}
\newcommand{\eeqa}{\end{eqnarray}}
\DeclareMathOperator{\sinc}{sinc}

\def\figureautorefname~#1\null{Fig.\,#1\null}
\def\tableautorefname~#1\null{Tab.\,#1\null}

\def\equationautorefname~#1\null{Eq.\,(#1)\null}

\RequirePackage[normalem]{ulem}

\crefname{section}{Sec.}{Secs.}
\crefname{figure}{Fig.}{Figs.}
\crefname{equation}{Eq.}{Eqs.}
\crefname{appendix}{Appendix}{Appendices}
\crefname{table}{Table}{Tables}

\def\CERN{Theoretical Physics Department, CERN, Esplande des Particules, 1211 Geneva 23, Switzerland}
\def\Cambridge{DAMTP, University of Cambridge, Wilberforce Road, Cambridge, UK}
\def\TAMU{Department of Physics and Astronomy, Mitchell Institute for Fundamental Physics and Astronomy, Texas A\&M University, College Station, TX 77843, USA}

\begin{document}

\preprint{CERN-TH-2023-195, MI-HET-818}

\title{
Broad Sterile Neutrinos \& the Reactor/Gallium Tension
}

\author{Hannah Banks}
\email{hmb61@cam.ac.uk}
\affiliation{\Cambridge}
\affiliation{\CERN}

\author{Kevin J. Kelly}
\email{kjkelly@tamu.edu}
\affiliation{\TAMU}

\author{Matthew McCullough}
\email{matthew.mccullough@cern.ch}
\affiliation{\CERN}

\author{Tao Zhou}
\email{taozhou@tamu.edu}
\affiliation{\TAMU}

\date{\today}

\begin{abstract}
Significant evidence exists for the apparent disappearance of electron-type neutrinos in radioactive source experiments. Yet, interpreted within the standard `3+1 sterile neutrino scenario', precision short-baseline measurements of electron antineutrinos from nuclear reactors strongly disagree with these results. Recently, it has been demonstrated that allowing for a finite wavepacket size for the reactor neutrinos can ameliorate such a tension, however the smallness of the required wavepackets is a subject of intense debate. In this work, we demonstrate that a `broad' sterile neutrino may relax this tension in much the same way.  Such a phenomenological possibility can arise in plausible hidden sector scenarios, such as a clockwork-style sector, for which we provide a concrete microscopic model.
\end{abstract}

\maketitle
\flushbottom

\section{Introduction}\label{sec:Introduction}
The number of neutrinos -- fundamental, neutral fermions of the Standard Model (SM) -- that exist has long been studied and debated. With observations of the $Z$-boson decay width, it has been determined that any such particles beyond the three expected to match the number of charged-lepton flavors (electron, muon, tau) must be `sterile', i.e.\ uncharged under the SM gauge group. As more has been learned about neutrino masses, mixing, and oscillation in the past twenty-plus years, several perplexing experimental results have bubbled up and persisted, potentially providing evidence for additional sterile neutrino states. To date, some of the strongest evidence comes from measurements of electron-flavor neutrinos from radioactive sources interacting with detectors constructed of gallium, wherein the observed interaction rate is significantly smaller than the expected rate in the standard, three-neutrino framework~\cite{GALLEX:1992gcp,SAGE:1994ctc,Barinov:2021asz, Barinov:2021mjj,Barinov:2022wfh, Giunti:2022btk}. However, when interpreted in the simplest `3+1-neutrino' framework, such apparent disappearance is inconsistent with both observations from solar neutrinos and measurements of reactor-antineutrino interaction rates at short baselines~\cite{Giunti:2021kab,Berryman:2021yan}.

Various solutions to this dilemma have been proposed in hope of explaining the apparent inconsistency between the different results, utilizing various mechanisms both within and beyond the SM~\cite{Brdar:2023cms,Arguelles:2022bvt,Hardin:2022muu,Farzan:2023fqa}. One of the most phenomenologically successful ideas for relaxing this tension comes from proposing that the neutrino wave-packet size is finite and (relatively) small, such that neutrinos propagating from nuclear reactors have time to significantly decohere on the length scales of interest in such experiments (e.g.\ NEOS~\cite{NEOS:2016wee} and PROSPECT~\cite{PROSPECT:2020sxr}), but not over the radioactive source ones (e.g.\ BEST~\cite{Barinov:2022wfh}). The origin of the wavepacket size required for this scenario  is unclear from a quantum-mechanical perspective~\cite{Akhmedov:2022bjs,Jones:2022hme,Smirnov:2022aab} but will nevertheless be tested by the upcoming JUNO medium-baseline reactor-antineutrino experiment~\cite{deGouvea:2020hfl,deGouvea:2021uvg}. In the standard 3+1 scenario, all new effects scale with baseline length $L$ divided by neutrino energy $E$ (where BEST probes $L \lesssim 1$ m and $E \approx 750$ keV, NEOS $L \approx 20$ m and $E \approx 5$ MeV, etc.). This approach relaxes the tension by modifying the expected dependence such that new oscillations exist coherently (for some fixed $L/E$) at small baseline lengths but are dampened out at larger distances.

In this work, we study an alternative, related phenomenological ansatz - that the sterile neutrino that exists beyond the standard three does not occupy a single, definite mass state.  Such a situation may be realized, for example, on the microscopic level, by UV-complete models with a band of multiple closely spaced mass eigenstates. 
This echoes some of the spirit of models involving large extra dimension constructions which also give rise to a band of states \cite{Machado:2011kt,Carena:2017qhd,Forero:2022skg}.  Whereas such models remain in significant tension with reactor rate measurements this is not the case for our approach, where the  overall mass scale and spectral distribution of the new states are, in principle, unrelated. To examine the potential impact of such a class of mass spectra in a general fashion, we take, as a convenient phenomenological ansatz, the mass spectrum attributed to the  fourth sterile state to comprise a top hat function. We dub this scenario, which we demonstrate is capable of significantly ameliorating the aforementioned tension, as a `broad sterile neutrino.' This approach parallels the studies of Ref.~\cite{Banks:2022gwq} in which deviations from the canonical 3-neutrino oscillation probabilities induced by the `broadening' of these states were systematically investigated. 

The remainder of this work is organized as follows.  In \cref{sec:Pheno} we introduce the phenomenological model deployed in our investigations, before presenting the results of our analysis on experimental neutrino data in  \cref{sec:Results}. Whilst the mass spectrum adopted in this analysis is not intended to represent a concrete UV-complete model, we emphasize that it constitutes a convenient means by which to examine the potential of more complex mass spectra to relax the tension between the different experiments, showing that significant improvement can be achieved. In ~\cref{sec:Models} we then offer an example of a specific model in which such a broadened mass spectrum arises. We discuss its relation to the phenomenological ansatz utilized in our analysis in addition to commenting on its potential cosmological implications. Finally, in \cref{sec:Conclusions},  some concluding remarks are provided.

\section{Experimental Landscape \& Phenomenological Approach}\label{sec:Pheno}
\begin{figure}[ht]
    \centering
    \includegraphics[width = \linewidth]{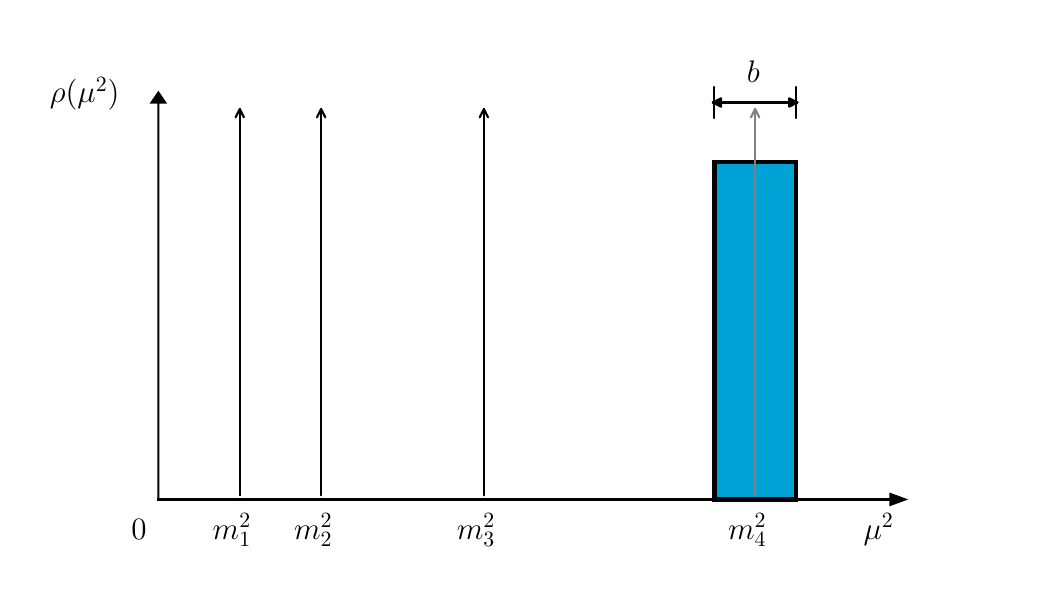}
    \caption{Sketch of spectral functions composed of 3 delta function (black arrows) and 1 top-hat function (blue box) as defined in \cref{eq: mass_spectrum}}
    \label{fig:mass_spectrum}
\end{figure}

Oscillations among the three, light, SM-like neutrinos have been measured to remarkable precision~\cite{Esteban:2020cvm,deSalas:2020pgw,Capozzi:2021fjo}, leading to a coherent three-flavor mixing paradigm. However, some other results may point to one or more additional `sterile' neutrino states around the eV scale. Following Refs.~\cite{Banks:2020gpu,Banks:2022gwq}, we allow for non-standard neutrino spectral densities $\rho(\mu^2)$ in the K{\"a}ll{\'e}n-Lehmann representation. Since the spectral densities of $m_{i}$, $1 \leq i \leq 3$, are well constrained~\cite{Banks:2022gwq}, we focus on the scenario in which the fourth (mostly-sterile) mass-eigenstate can be modeled as a state with a central mass value at $\mu^2 = m_4^2$ with some finite breadth $b$.   The spectral densities are thus
\begin{equation} \label{eq: mass_spectrum}
\rho_{e e}\left(\mu^{2}\right)=\left\{\begin{array}{lll}

|U_{e 1}|^2\delta(\mu^2-m_1^2) & , & \mu^2=m_1^2 \\ 

|U_{e 2}|^2\delta(\mu^2-m_2^2) & , & \mu^2=m_2^2 \\ 

|U_{e 3}|^2\delta(\mu^2-m_3^2) & , & \mu^2=m_3^2 \\ 

\frac{1}{b}\left|U_{e 4}\right|^{2} ,~~ m_4^2-\frac{b}{2} \leq &\mu^2& \leq m_4^2+\frac{b}{2} \end{array}\right\},
\end{equation}
where $U_{ei}$ represent the elements of the (extended $4\times4$) leptonic mixing matrix. We use the standard parameterization in which $U_{e4} = \sin\theta_{14}$. This spectral density is illustrated in \cref{fig:mass_spectrum}. Naturally, we will only be sensitive to this scenario with experiments that have access to the energy/length scales dictated by $m_4^2$ and/or $b$.

The amplitude for $\nu_e \to \nu_e$ transitions (relevant for reactor antineutrino oscillations) can be expressed as 
\begin{align} \label{eq:amplitude}
i\mathcal{A}_{ee} &= \int_{0}^{\infty} d\mu^2 e^{-i\frac{\mu^2 L}{2E}} \rho_{ee}\left(\mu^2\right) \nonumber\\
&= \sinc\left(\frac{bL}{4E}\right)|U_{e4}|^2 e^{-\frac{iLm_4^2}{2E}}+\sum_{i=1}^3 |U_{ei}|^2 e^{-\frac{iLm_i^2}{2E}}.
\end{align}
In the limit that $\Delta m_{41}^2 \gg |\Delta m_{31}^2|,\ \Delta m_{21}^2$ and that the phases associated with $\Delta m_{21}^2$ and $\Delta m_{31}^2$ have yet to develop, the oscillation probability $P_{ee} \equiv |\mathcal{A}_{ee}|^2$ can be approximated as
\begin{align}\label{eq:probability}
P_{ee} &\simeq \left(1 + \left(\sinc{\left(\frac{b L}{4E}\right)}-1\right)|U_{e4}|^2\right)^2 \\
&-4|U_{e4}|^2 \left(1 - |U_{e4}|^2\right)\sin^2{\left(\frac{\Delta m_{41}^2 L}{4E}\right)} \sinc{\left(\frac{b L}{4E}\right)}.\nonumber 
\end{align}
Here we have defined $\Delta m_{ij}^2 \equiv m_i^2-m_j^2$. The sinc term appearing in \cref{eq:probability} provides an energy-dependent overall normalization as well as an energy-dependent modification of the effective mixing angle between the sterile and active states. The dominant effect of the breadth $b$ in this respect is to replace $\sin^2\left(2\theta_{ee}\right) = 4|U_{e4}|^2(1-|U_{e4}|^2)$ with $\sin^2\left(2\theta_{ee}^{\rm eff.}\right)(E)$, where
\begin{equation}
\sin^2\left(2\theta_{ee}^{\rm eff.}\right)(E) = \sin^2\left(2\theta_{ee}\right) \sinc{\left(\frac{b L}{4E}\right)}.
\label{eq:effmix}
\end{equation}
This effect mirrors the exponential suppression present in the decoherence model studied in Refs.~\cite{Arguelles:2022bvt,Hardin:2022muu} and the decay model in Ref.~\cite{Hardin:2022muu}, particularly when $L/E$ is large (relative to the size of $b$). For this reason, we expect the broad sterile neutrino scenario introduced here to be able to achieve a similar impact on neutrino data as the decoherence and decay scenarios.

\begin{figure}[ht]
    \centering
    \includegraphics[width = \linewidth]{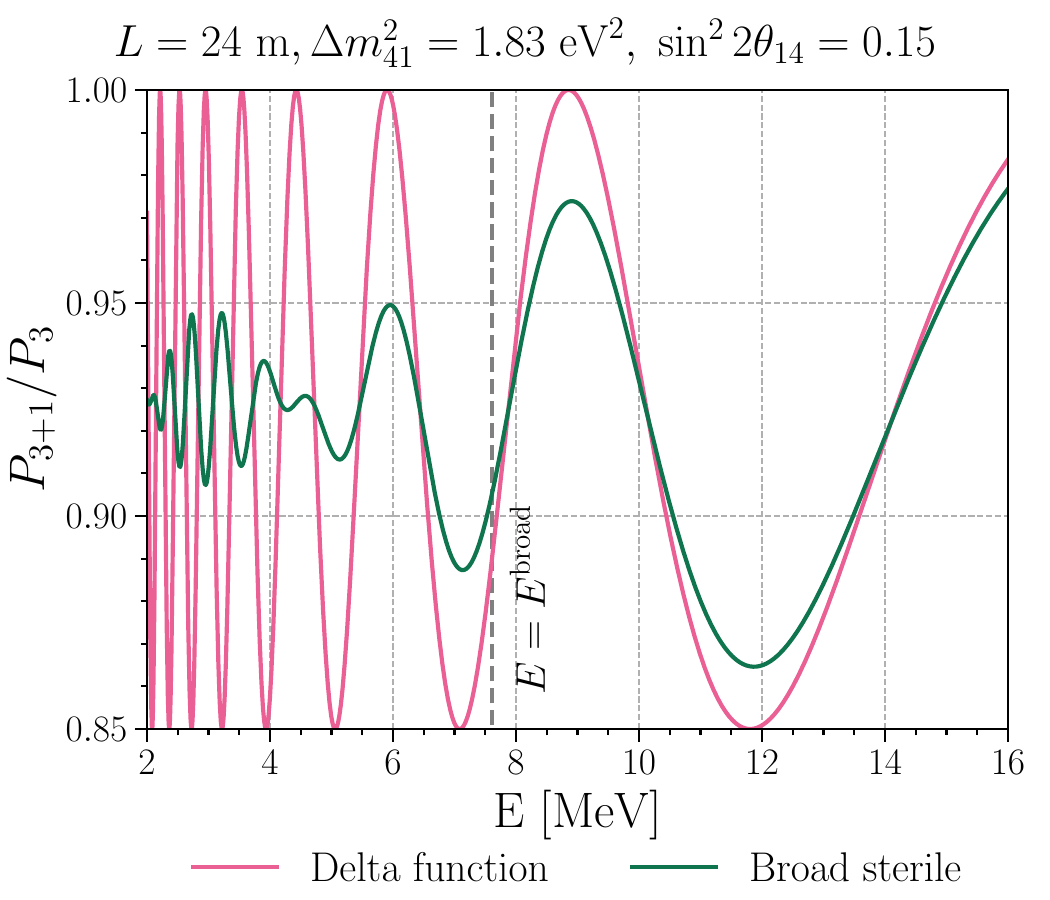}
    \caption{Demonstration of the damped oscillations from the broad sterile neutrino. We plot the ratio of the oscillation probabilities between the 3+1 sterile neutrino model and the three-neutrino model as a function of energy, assuming a baseline of 24 m and $\Delta m_{41}^2 = 1.83$ eV$^2$. The pink line corresponds to the vanilla $3+1$ scenario, whereas the green line assumes $\tilde{b} = b/\Delta m_{41}^2 = 0.24$ such that the low-energy damping effect is apparent.}
    \label{fig:probability_ratio}
\end{figure}

We demonstrate the impact of nonzero $b$ on the oscillation probability in~\cref{fig:probability_ratio}, akin to the presentations in Refs.~\cite{Arguelles:2022bvt,Hardin:2022muu}. Concretely, in the broad sterile neutrino scenario (green) we assume $L = 24$ m (as in NEOS), $\Delta m_{41}^2 = 1.83$ eV$^2$, $\sin^2 2\theta_{14} = 0.15$, and $b = 0.24 \Delta m_{41}^2$. The pink line demonstrates the oscillation probability, relative to the three-flavor case, with $b = 0$. Similar to the above references which study the decohering and decaying sterile-neutrino scenarios, we find a dampening of the new oscillations at relatively low neutrino energies, relative to the vanilla $3+1$ case. In contrast, however, the $\sinc$ function presented here causes a `rebounding' of the oscillation probability, which could allow for distinctions between the various scenarios. The effect of the $\sinc$ term is most evident below the energy we have labeled `$E = E^{\rm broad}$', where the $\sinc$ function begins to contribute noticeably to the oscillation probability.

\section{Analysis \& Results}\label{sec:Results}
With this formalism in mind, we turn to the impact of such a broad sterile neutrino on neutrino oscillation data. For clarity, we assume a normal mass ordering and that $m_4 \gg m_1$, so that $\Delta m_{41}^2 \approx m_4^2$. We introduce a (dimensionless) reduced breadth parameter $\tilde{b} \equiv b/m_4^2 \approx b/\Delta m_{41}^2$ such that $\tilde{b} < 2$. We include this scenario into the framework presented in Ref.~\cite{Arguelles:2022bvt} (code available at~\cite{githubGitHubHarvardNeutrinoDayaBaySterileDecoherence}) which allows us to analyze the Daya Bay~\cite{DayaBay:2016qvc,DayaBay:2016ggj}, NEOS~\cite{NEOS:2016wee}, PROSPECT~\cite{PROSPECT:2020sxr}, and BEST~\cite{Barinov:2021asz} experiments either individually or simultaneously.

\begin{figure}[ht]
    \centering
    \includegraphics[width = \linewidth]{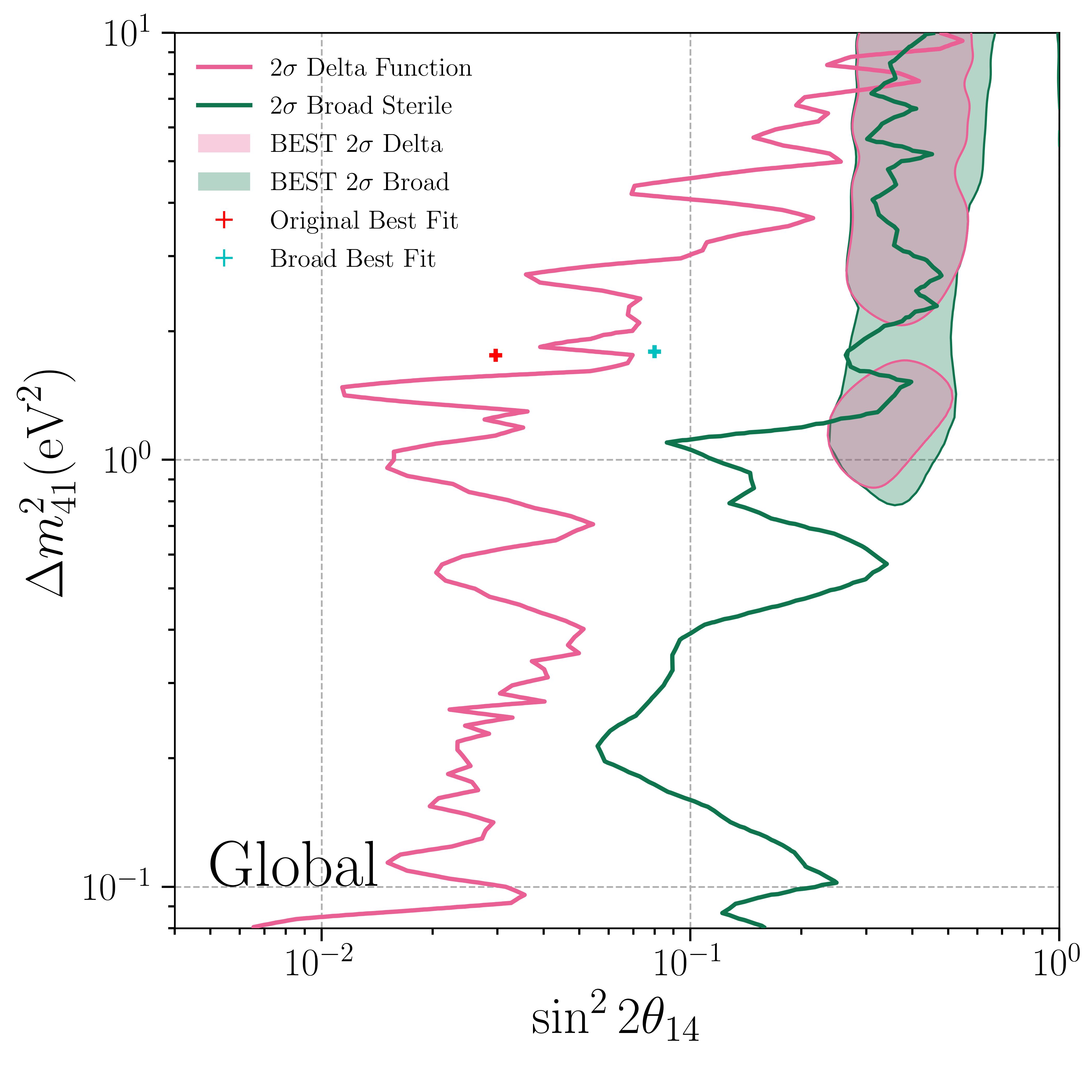}
    \caption{Joint constraint from reactor antineutrino experiments (solid lines) vs. the preferred region by BEST (shaded) at $2\sigma$ CL. Pink line/regions correspond to the standard $3+1$ scenario, where green allows for nonzero sterile neutrino breadth $b$ (where we have marginalized over this parameter in the analysis). Crosses indicate the best-fit point of the reactor analyses in each scenario -- see text for more detail. 
    \label{fig:ContourComparison}}
\end{figure}

\cref{fig:ContourComparison} shows the main results of this analysis. The pink line and shaded regions respectively show the constraints from reactor experiments and the parameter space preferred by the BEST experiment, under the standard $3+1$ sterile neutrino scenario. Significant tension is evident between these results. Considering nonzero breadth, we allow $\tilde{b}$ to vary and find that (a) the parameter space preferred by BEST does not change substantially (green shaded region) and (b) the constraints by reactor experiments on $\sin^2 2\theta_{14}$ as a function of $\Delta m_{41}^2$ relax considerably. We present these results at $2\sigma$ confidence level and see that, while the reactor analysis' best-fit point (green cross) is far outside the preferred region of BEST, there is significant overlap between the datasets for relatively large mixing angles. Similarly to Ref.~\cite{Arguelles:2022bvt}, we present the reactor antineutrino results as a $2\sigma$ CL ($\Delta \chi^2 = 6.18$) upper limit with respect to the \textit{null hypothesis}, rather than the best-fit point, which allows for a slightly more conservative limit and guards against statistical fluctuations leading to (potentially) overstated significance for sterile neutrino evidence. See, e.g.\ Refs.~\cite{Giunti:2021kab,Berryman:2021yan,Giunti:2022btk} for further discussion. The reactor-only best-fit point moves from ($\sin^2 2\theta_{14},~\Delta m_{41}^2$) = ($0.03$, $1.76$ eV$^2$) to ($0.09$, $1.73$ eV$^2$) when $b\neq 0$ is allowed. 

Not drawn in~\cref{fig:ContourComparison} is the constraint on such a scenario from solar neutrino measurements~\cite{Goldhagen:2021kxe,Berryman:2021yan}, which disfavors $\sin^2 2\theta_{14} \gtrsim 0.15$ ($0.23$) at $2\sigma$ ($3\sigma$) CL in the $3+1$ scenario. Like with the decoherence scenario~\cite{Arguelles:2022bvt}, we expect that nonzero $b$ will not significantly relax such a constraint and so this tension remains.

\begin{figure}[ht]
    \centering
    \includegraphics[width = \linewidth]{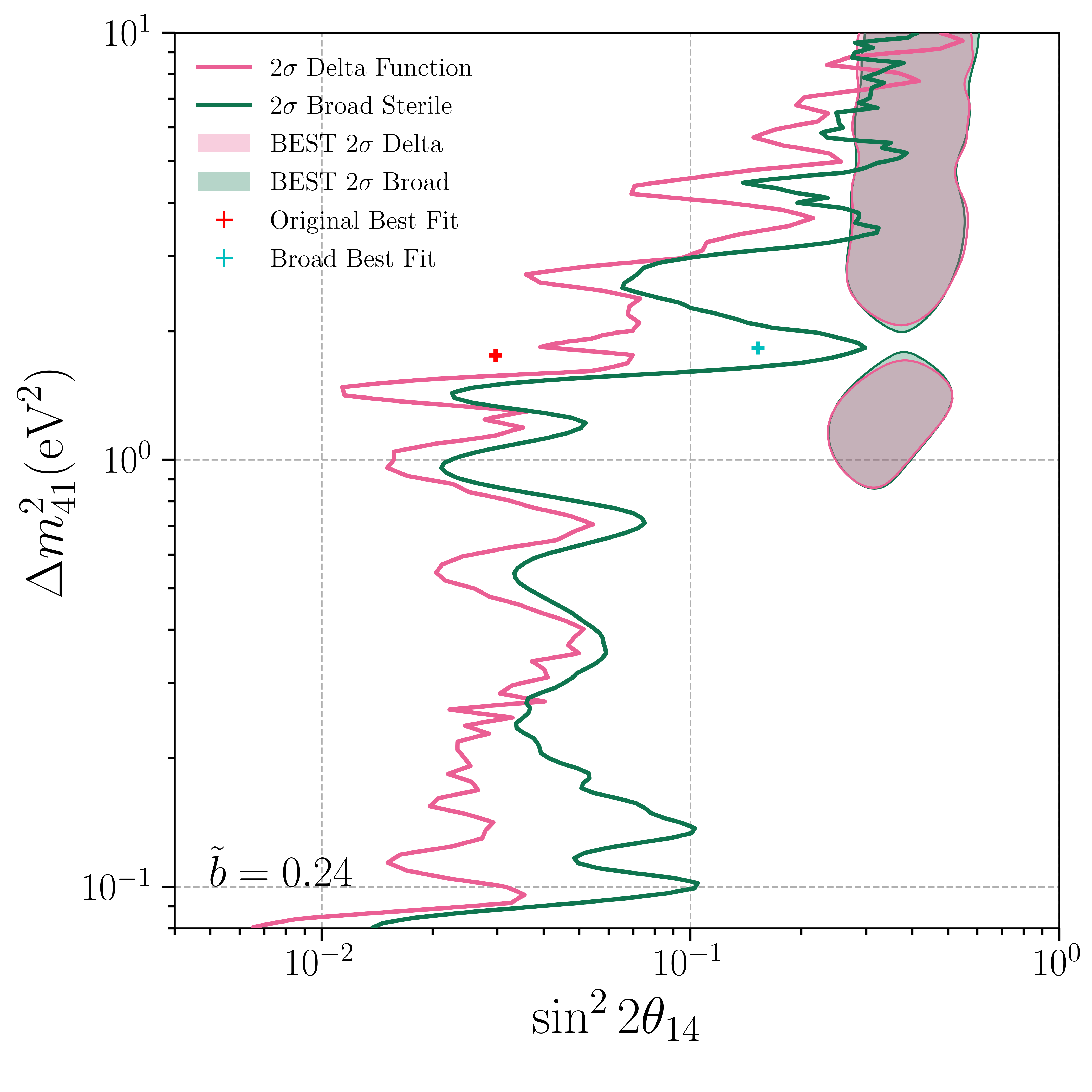}
    \caption{Same as~\cref{fig:ContourComparison} but with $\tilde{b} = 0.24$ fixed instead of marginalized.
    \label{fig:ContourComparison_b=0.24}}
\end{figure}

In contrast to the marginalized-$b$ \cref{fig:ContourComparison}, in~\cref{fig:ContourComparison_b=0.24} the data is analyzed with fixed $\tilde{b} = 0.24$. Because we have fixed this parameter, the green constraints are not as relaxed (compared to the $b = 0$ ones), but agreement does start to appear between reactors and BEST for $\Delta m_{41}^2 \approx 5$ eV$^2$.
Overall, the newfound agreement between the reactor constraints and the BEST preferred region arises because the breadth dampens the (expected) oscillations at (primarily) NEOS and Daya Bay, but does not have a large impact on the relatively small BEST scales.

\begin{figure}[ht]
    \centering
    \includegraphics[width = \linewidth]{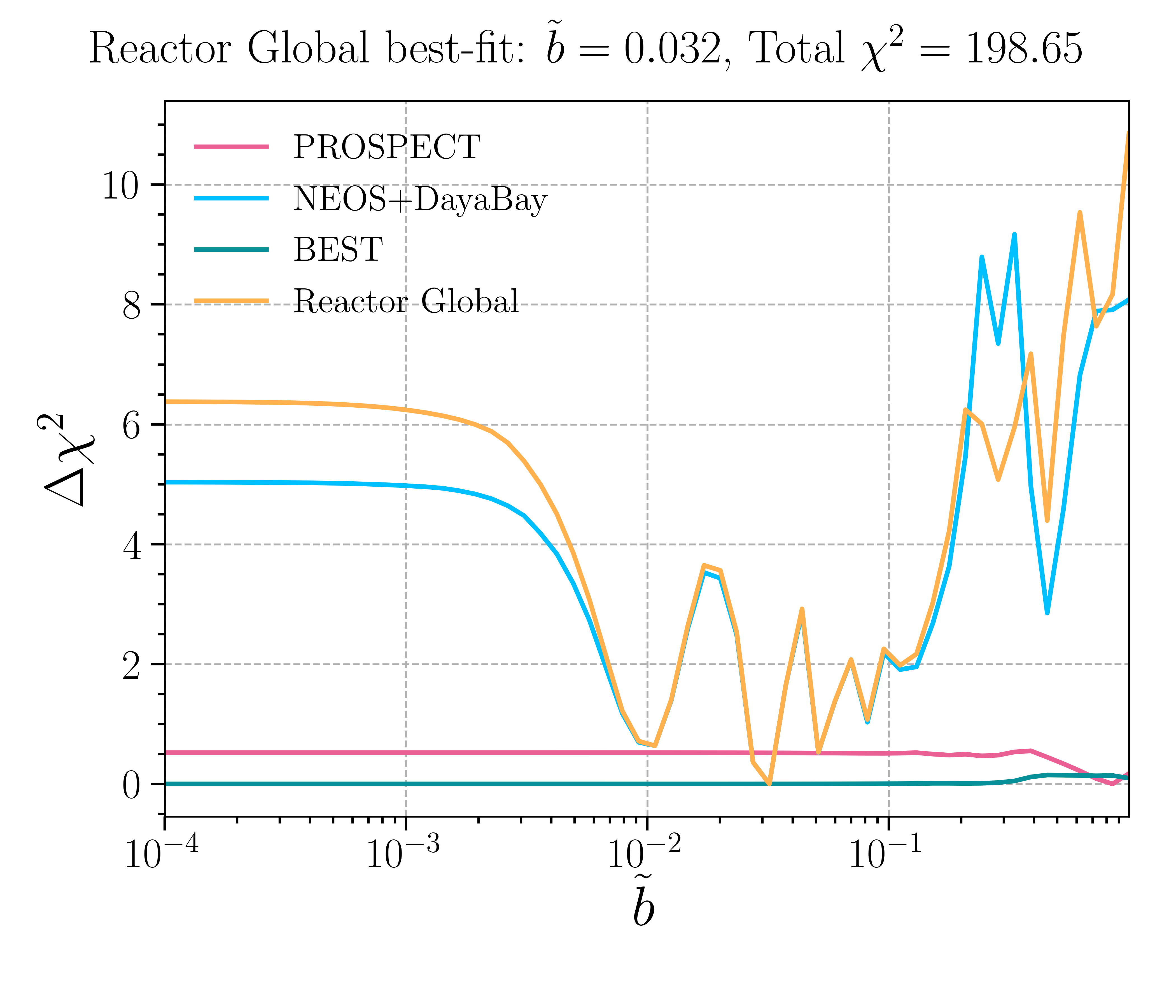}
    \caption{$\Delta \chi^2$ vs. $\tilde{b}$ relative to its minimum value, marginalizing over $\Delta m_{41}^2$ and mixing for various experiments. The best-fit mass and mixing are $\Delta m_{41}^2=1.73$eV$^2$, $\sin^2(2\theta_{14})=0.09$.}
    \label{fig:Chi2b}
\end{figure}

Finally, we also are interested in the rough value of $\tilde{b}$ preferred in such an analysis. \cref{fig:Chi2b} demonstrates the test-statistic $\Delta \chi^2$ (relative to the best-fit) as a function of $\tilde{b}$, marginalized over the mixing angle and mass-squared splitting. The best fit lies at $\tilde{b} = 0.032$ ($b = 0.055$ eV$^2$), and improves the global reactor fit by approximately $6$ units of $\chi^2$. We summarize the quality of this fit and its improvement over the delta function $3+1$ scenario in~\cref{tab:chi2}.
Effectively, the tension is reduced when $b\neq 0 $ is allowed because it allows for a smaller effective mixing angle at the reactor experiments (operating at larger $L/E$ than the source ones (at lower $L/E$), where this effective mixing angle is suppressed by $b$. We find similar improvement to this data to the decohering sterile neutrino hypothesis. We leave the statistical significance of this improvement to future work, but briefly highlight the prospects of disentangling these two scenarios. The decohering sterile neutrino hypothesis (as well as the decaying one) predicts the oscillation amplitudes to go to zero for sufficiently large $L/E$, whereas our broadened scenario predicts this to ``rebound.'' In principle, precise enough low-energy reactor antineutrino data could discern the mechanism at play.

\begin{table}[htbp]
  \centering
  \begin{tabular}{l c c}
    \hline
    \textbf{Experiments} & \textbf{$\chi^2/\textrm{dof}$} ($b\neq 0$) & \textbf{$\chi^2/\textrm{dof}$} ($b=0$)\\
    \hline
    PROSPECT & 125.7/157 & 126.1/158\\
    NEOS+DayaBay & 68.9/137 & 73.9/138\\
    Reactor Global & 198.7/297 & 205.0/298\\
    \hline
  \end{tabular}
  \caption{Best fit $\chi^2/\textrm{dof}$ for various experiments.}
  \label{tab:chi2}
\end{table}

\section{Complete Model}\label{sec:Models}
It remains to present a concrete microscopic theory which could lead to such a broad sterile neutrino.  Our goal here is not to build a model that exactly fits the data analysis under the phenomenological approach presented in~\cref{sec:Pheno,sec:Results},  but to provide a demonstrative example of a scenario which shares the desired phenomenological features required to reduce the experimental tension. We will assume that the active SM neutrinos are Majorana, such that at the energy scales of interest their masses are generated via the usual lepton-number-violating Weinberg operator
\begin{equation}
\mathcal{L}_\nu = \frac{1}{\Lambda} \lambda_{ij} (L_i H) (L_j H) ~~.
\label{eq:weinberg}
\end{equation}
The generated neutrino mass matrix is assumed to be diagonalized by the usual PMNS matrix, such that $\nu_\alpha = U_{\alpha i} \nu_i$. The main novel ingredient in this construction is that these states will mix with a band of sterile neutrinos with a mass spectrum centred around a value $M$ and with an effective mass-squared width $b$. 

There are many scenarios that could give rise to such a band of sterile states, particularly if a hidden sector involved a strongly coupled composite sector.  Consider, for instance, a QCD-like sector with two light quarks and no electroweak group and keep in mind throughout, for intuition, the QCD sector of the Standard Model.  At low masses there are no composite fermions, due to the mass gap.  Just above the mass gap one
would have the equivalent of the proton, neutron, and then above that a number of composite fermionic states. Continuing to higher masses, once well above the scale of strong coupling there are no new fermionic resonances as individual quark production channels open. Thus such a QCD-like sector would give rise to a band of fermionic resonances in the region of the strong coupling scale which are sterile to us SM observers.  The fact that such a band of fermionic states has already arisen in the visible sector lends weight to the possibility that it could also do so in a hidden sector. 
 
For concreteness we may model such a sector by employing a `clockwork-inspired' construction as per Refs. \cite{Choi:2015fiu,Kaplan:2015fuy,Giudice:2016yja}.  In this model one has a ring of sterile Majorana fermions `$S$' with site translation-invariant clockwork-inspired masses
\begin{equation}
\mathcal{L} \supset \frac{M}{2 q^2} \left[ \sum_{\text{Ring}} S_j \left( (1+q^2) S_j-2 q \left( S_{j-1} + S_{j+1}\right)\right) \right] ~.
\end{equation}
The mass matrix is
\begin{equation}
M_S = \frac{M}{q^2}
\begin{pmatrix}
1+q^2 & -q & 0 & \cdots &  & -q \cr
-q & 1+q^2 & -q & \cdots &  & 0 \cr
0 & -q & 1+q^2 & \cdots & & 0 \cr
\vdots & \vdots & \vdots & \ddots & &\vdots \cr
 0 & 0 & 0 &\cdots & 1+q^2 & -q \cr
 -q & 0 & 0 &\cdots & -q & 1+q^2
\end{pmatrix} \ ~~,
\label{psimass}
 \end{equation}
with eigenvalues
\begin{equation}
m_J = \frac{M}{q^2} \left(1+q^2 - 2 q \cos \left(\frac{2 \pi J}{N} \right) \right) ~~.
\label{eq:masses}
\end{equation}
For $q\gg1$ we thus have a band of states centered at mass $M$ with fractional width scaling as $\propto 1/q$.  The rotation matrix for the mass matrix is
\begin{equation}
R^{\rm CW}_{JK}  =  \frac{\cos\left( \frac{2 \pi JK}{N} \right) + \sin\left( \frac{2 \pi JK}{N} \right)}{\sqrt{N }} ~~.
\end{equation}
Thus far the SM and hidden sectors are decoupled and may be independently diagonalized by the matrices $U$ and $R^{\rm CW}$ respectively.

In linking the two sectors we assume that the only interaction between the active and sterile sectors is through the neutrino portal
\begin{equation}
\mathcal{L}_{\nu N} = \sum_{iI} \alpha_{Ii} S_I  L_i H ~~,
\label{eq:neutint}
\end{equation}
where $\alpha$ is an $N \times 3$ matrix, where $N$ is the number of sterile neutrino mass eigenstates.

The combined mass matrix in the weak eigenstate basis is thus
\begin{equation}
M = 
\begin{pmatrix}
M_\nu & \alpha_*   \cr
\alpha_*^T & M_S
\end{pmatrix} \ ~~,
 \end{equation}
 where $\alpha_* = \alpha v / \sqrt{2}$ and $v$ denotes the Higgs VEV. It is this matrix which must, provided a given  $\alpha$, be diagonalized to reveal the physical neutrino mass and mixing matrices required to compute the oscillation probabilities. 

With the model thus outlined, we seek to investigate its correspondence to the effective `broad' description deployed in our above analysis. We do this by seeking a  set of model parameters for which the $\nu_e \rightarrow \nu_e$ oscillation probability as a function of $L/E$ approximates that predicted by our broad ansatz for the parameters which best fit the experimental data ($\tilde{b}$ = 0.24, $~\sin^2 2\theta_{14} = 0.15, ~\Delta m_{41}^2 = 1.83\ \textnormal{eV}^2$), whilst remaining consistent with the usual mass-squared splittings and mixings between the active neutrinos as determined experimentally. 

In this latter respect, for both simplicity and demonstrative purposes, we work in the `alignment' limit in which the matrix $O$ required to diagonalize $M$ can be factorized to the form \begin{equation}
O = 
\begin{pmatrix}
\cos (\theta) U & \sin (\theta) T   \cr
-\sin (\theta) T^T & \cos (\theta) \widetilde{R}
\end{pmatrix} \ ~~,
\end{equation} where, $U$ is the usual PMNS matrix, $\widetilde{R}$ is an $N \times N$ rotation matrix and $T$ is a $3 \times N$ matrix defined according to
\begin{equation}
T_{ij} = \begin{cases}
			1, & \text{if $i$ = 1}\\
            0, & \text{if $i$ $\neq$ 1}
		 \end{cases}~~,
\end{equation} $\forall j \in {1,...,N}$. 

By definition one has that 
\begin{equation}
M = O M^D O^T ~~,
\label{tmass}
\end{equation}
where
\begin{equation}
M^D = \begin{pmatrix}
\widetilde{M}^D_{\nu} & 0  \cr
0 & \widetilde{M}^D_S~~, 
\end{pmatrix} \
\end{equation}
and $\widetilde{M}^D_{\nu}$ and $\widetilde{M}^D_{S}$ are respectively $3 \times 3 $ and $N \times N$ diagonal matrices containing the physical mass eigenvalues of the active and sterile neutrinos. 

For the purpose of comparison, we construct a matrix $M$ according to~\cref{tmass}, setting $\widetilde{R}$ to $R^{\rm CW}$, $\widetilde{M}^D_{\nu}$ to diag(0,$\sqrt{\Delta m^2_{21}}$,$\sqrt{\Delta m^2_{31}}$) and $\widetilde{M}^D_S$ to $(R^{\rm CW})^T M_S R^{\rm CW}$. We then diagonalize this numerically to obtain the physical mass, $M_{n}^D$ and mixing, $O_{n}$ matrices.

We emphasize that whilst the eigenvalues that we obtain post diagonalization will not equal the `guesses' used as input in our starting choices of $\widetilde{M}^D_{\nu}$ and $\widetilde{M}^D_S$  exactly, due the form of $O$, the differences are small. Similarly, the upper $3 \times 3$ sub-matrix of $O_n$ that encodes the mixing between the active states, should, by construction, approximately equal the input PMNS matrix $U$. 

We compute the $\nu_e \rightarrow \nu_e$ oscillation probability from the numerically obtained mass and mixing matrices according to  
\begin{equation}
P(\nu_e \rightarrow \nu_e) = \left|\sum_{j = 1}^{N+3} |(O_{n})_{1j}|^2 e^{-\frac{iLm_j^2}{2E}} \right|^2~,
\end{equation}
where $m_j \equiv (M^D_{n})_{jj}$.

With these choices in hand, there remain   4 free parameters - $N$, $M$, $q$ and $\theta$. For the purpose of example we fix $N = 40$ and perform a least squares fit to the `best fit' broad  probability distribution as a function of $L/E$ over the range 1-10 m/MeV fitting for $M$, $q$ and $\theta$.

\begin{figure}[ht]
    \centering
    \includegraphics[width = \linewidth]{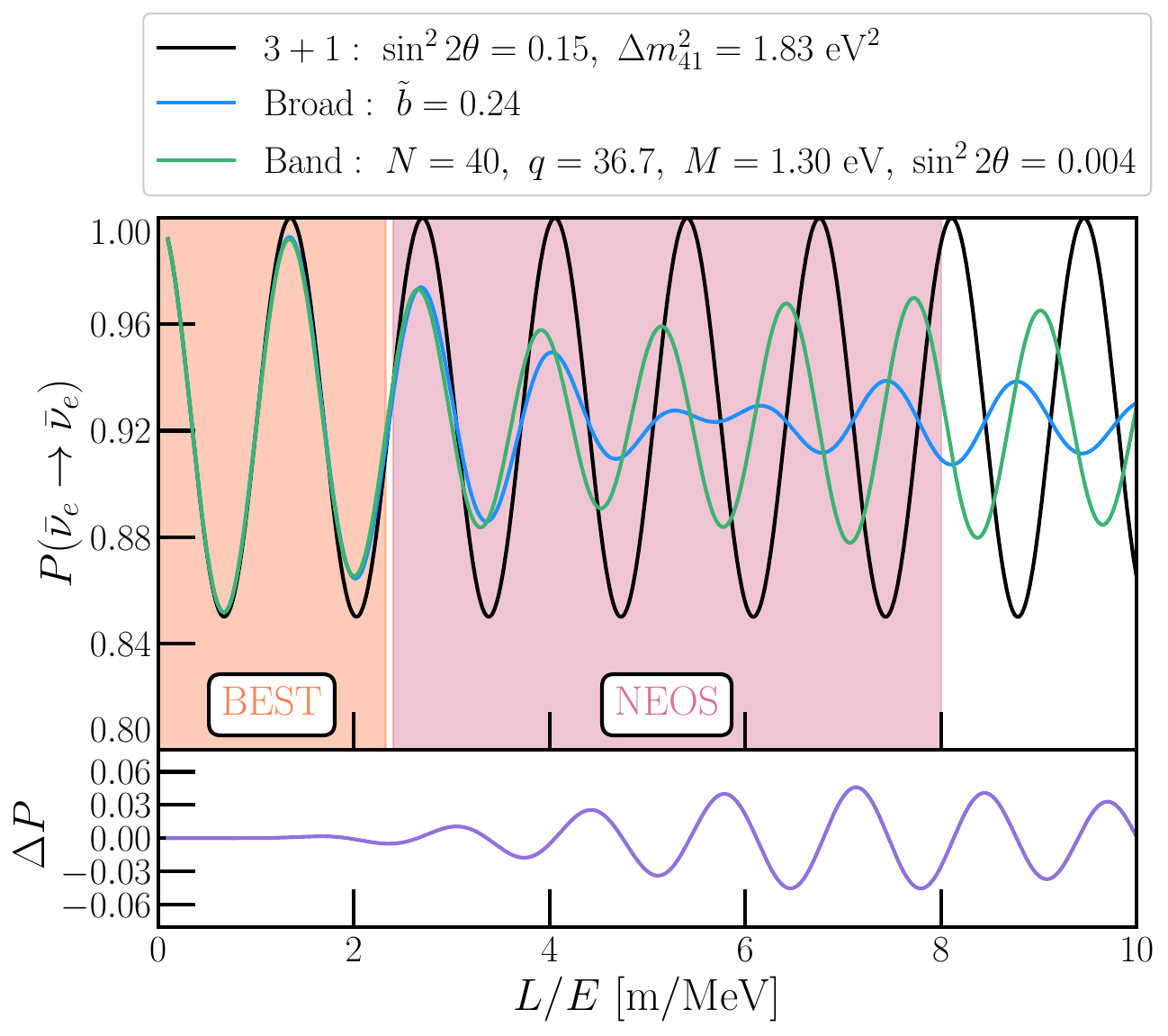}
    \caption{Comparisons of $\bar\nu_e \to \bar\nu_e$ oscillation probabilities as a function of $L/E$ determined from the standard 3+1 scenario (black), the same with a broadened fourth state (blue), and from the `best-fit' band scenario (green) obtained by a least-squares fit. Parameters for each are given in the legend. Colored shaded regions in the top panel indicate portions of $L/E$ to which BEST and NEOS are most sensitive.}  
    \label{fig:bb}
\end{figure}

The agreement between the probability distributions as a function of $L/E$ for the selected broad case and corresponding best-fit of the concrete band model is displayed in Fig.~\ref{fig:bb}. For reference, the standard 3+1 oscillation probability for the same $\theta_{14}$ and $\Delta m_{41}^2$ as used in the broad case is shown in black. The lower panel shows the residuals of the fit of the band model to the broad case. The concordance between the band model and broad ansatz is excellent at low $L/E$ (including the entirety of the domain measured at BEST) where the experiments are unable to resolve the different microscopic splittings within the band. As expected, the oscillation probability for the band model deviates from that of the broad ansatz at higher $L/E$ as the experiments gain sensitivity to the differences in the microphysical structure of their mass spectra. Crucially for the desired amelioration of tension however, we note that the example band model shares the favorable characteristic of the broad ansatz in that the amplitude of the oscillations in the $\bar\nu_e$ disappearance probability are damped at reactor experiments relative to the radioactive source ones.   

\begin{figure}[ht]
    \centering
    \includegraphics[width = \linewidth]{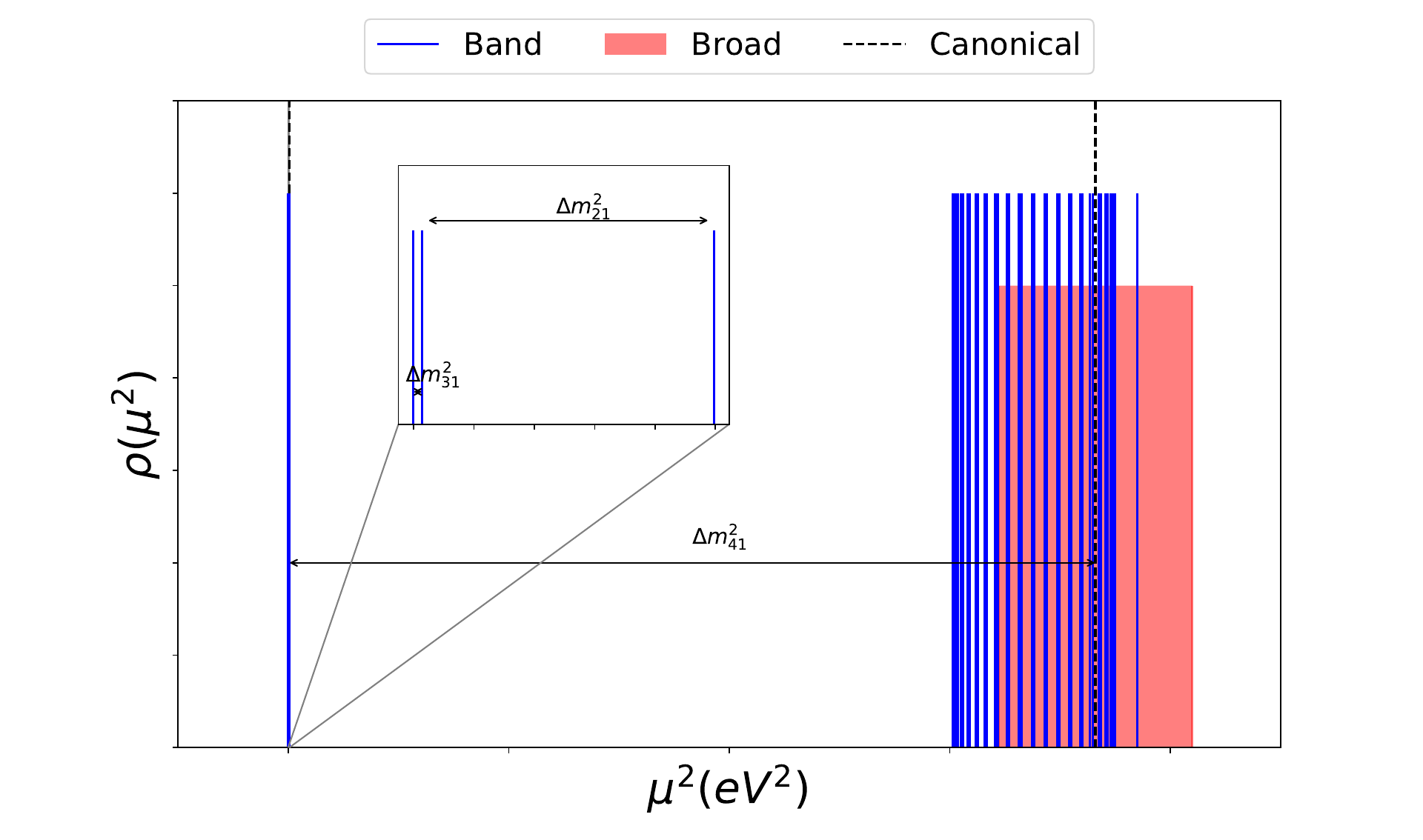}
    \caption{A plot of the spectral function $\rho(\mu^2)$ for the broad and the band cases, using the same parameters as in Fig.~\ref{fig:bb}. The vertical extent of the formally infinite $\delta$ functions corresponding to the band case are for illustrative purposes only.}
    \label{fig:ms}
\end{figure}

In Fig.~\ref{fig:ms} we plot the density of states for these models using the same parameters as detailed above. Corresponding to formally infinite delta functions, the vertical extent of the states for the band case are for illustrative purposes only. We note that the three lightest states in the band and broad case are (up to small corrections) coincident, and thus plotted here in blue only. 

Generalizing slightly we can now understand the connection between the mixing angles in the two approaches.  If an effective broad sterile neutrino arises microscopically as a result of a band of $N$ sterile neutrinos each with a mixing angle $\sim\theta$ with the active neutrinos then the total effective mixing angle of the broad sterile neutrino, $\theta_{\text{b}}$ will be approximately given by
\begin{equation}
\sin^2 \theta_{\text{b}} \approx N \sin^2 \theta ~~.
\end{equation}
For example, if we consider $N = 40$, then an effective mixing angle $\sin^2 2\theta_{\text{b}} \approx 0.15$ required by BEST may be generated by $\sin^2 2\theta \approx 0.004$, as observed in Fig.~\ref{fig:bb}.

\subsection*{Cosmology}
Cosmological production rates for a single sterile neutrino state with active-sterile mixing angle $\theta$ will scale proportional to $\sin^2 \theta$.  Thus, for a band of states the total abundance of produced sterile states will scale as $N \sin^2 \theta$.  However, we already saw that this is the same as $\propto \sin^2 \theta_{\text{b}}$, for the effective broad neutrino.  Thus for a single broad neutrino, no matter the underlying microscopic narrow states which give rise to the effective breadth, one expects to generate a similar cosmological abundance as would be generated for a single narrow sterile neutrino state with mixing angle $\theta_{\text{b}}$.  In other words, there is no special $N$-dependence which can significantly modify the scaling of cosmological production.

Sterile neutrinos have a rich and complex impact on the early universe and observational cosmology.  See, for instance, Ref.~\cite{Dasgupta:2021ies} for a recent review.  While a broad sterile scenario can render a gallium anomaly explanation consistent with reactor data, an additional source of tension comes from cosmological $N_{\text{eff}}$ determinations, from BBN, CMB and BAO observables (see also Refs.~\cite{Brdar:2023cms,Hagstotz:2020ukm}).  For the same mixing and mass a broad sterile neutrino will give a comparable $N_{\text{eff}}$ contribution as an additional narrow sterile neutrino, hence the scenarios proposed here will also be in tension to the same overall degree.

On the other hand, many of the proposed resolutions for ameliorating cosmological tensions in the case of standard sterile neutrinos can apply here -- see Ref.~\cite{Gerbino:2022nvz} for a recent review.  For instance, if the states comprising the band of sterile neutrinos can decay into an active neutrino and radiation, such as a new ultralight boson, on a fast enough timescale, then predictions can be brought into line with cosmological observations.

\section{Discussion \& Conclusions}\label{sec:Conclusions}
In this work we have reassessed a long-standing neutrino-physics tension between the null results of reactor antineutrino experiments and the putative positive signal in gallium source experiments. We have drawn on previous analyses which suggested that decoherence from small neutrino wavepackets is at play. Building on this suggestion, we have demonstrated that a `broad sterile neutrino' scenario, where the additional neutrino species does not occupy a single definite mass state, is qualitatively similar in relaxing the tension between these experimental results.

We have shown that this scenario provides a relatively good, albeit not perfect, fit to the data of interest. In doing so, we have identified the rough scale of the sterile neutrino breadth as a target for future experimental efforts. Such nonzero breadth could be discovered in future studies of reactor antineutrino experiments.

Finally, we have placed a significant emphasis on the fact that such a broad sterile neutrino could be realized in a variety of UV-complete, beyond-the-Standard-Model theories. The concrete model we have focused on resembles a clockwork scenario of new fermions, strongly coupled to each other and weakly coupled (via mixing) to the Standard Model neutrinos. We have shown that, for length and energy scales of interest, this model can be reasonably well described by the simple phenomenological approach we have used in fitting data.  It would be interesting to investigate if alternative microscopic models could improve the gallium/reactor tension even further, since the shape of the phenomenological model has not been optimized to this end.

Whether the evidence from gallium source experiments is \textit{bona fide} new physics remains to be seen, but, if it is, additional phenomena are required to resolve tension with reactor antineutrino measurements, as well as predictions regarding solar neutrinos and cosmology which lie beyond our main focus. Broad sterile neutrinos provide a novel and interesting class of potential solutions.
\\

\begin{acknowledgments}
We would like to thank Toni Bert{\'o}lez-Mart{\'i}nez for sharing his fitting codes and providing detailed clarification. We are grateful to Joachim Kopp for valuable discussions regarding this work. We also acknowledge Texas A\&M University High Performance Research Computing (HPRC) for providing computing resources. KJK and TZ acknowledge support from the United States DOE grant DE-SC0010813. HB acknowledges partial support from the STFC HEP Theory Consolidated grants ST/T000694/1 and ST/X000664/1 and thanks other members of the Cambridge Pheno Working Group for useful discussions.
\end{acknowledgments}

\bibliographystyle{JHEP}
\bibliography{refs}

\end{document}